\documentclass{article}
\usepackage{spconf}
\usepackage{amsmath, amssymb, amsthm, amsfonts}
\usepackage[hidelinks]{hyperref}
\usepackage{graphicx}
\usepackage{booktabs}
\usepackage{bm}
\usepackage{cite}
\usepackage{color}
\usepackage{mathdots}
\usepackage{optidef}
\usepackage{soul}
\usepackage[caption=false,font=normalsize,labelfont=sf,textfont=sf]{subfig}
\usepackage{orcidlink}

\usepackage{textcomp}

\DeclareMathOperator*{\argmin}{arg\,min}
\DeclareMathOperator*{\argmax}{arg\,max}

\DeclareMathOperator{\sinc}{\text{sinc}}
\DeclareMathOperator{\sinhc}{\text{sinhc}}

\newcommand{\transpose}{\top}

\newcommand{\real}{\mathbb{R}}

\newcommand{\naturals}{\mathbb{N}}
\newcommand{\unitsphere}{\mathbb{S}^2}

\DeclareMathOperator*{\minimize}{minimize}

\usepackage{tikz}
\usepackage{pgfplots}
\pgfplotsset{compat=newest}
\usetikzlibrary{calc, math, positioning, backgrounds, arrows, arrows.meta, bending, decorations.markings, decorations.text, plotmarks, patterns.meta, patterns, decorations.pathreplacing}
\usepgfplotslibrary{external}
\usepgfplotslibrary{groupplots}

\pgfplotsset{
	layers/my layer set/.define layer set={
		bg,
		main,
		foreground
	}{
	},
	set layers=my layer set,
}

\colorlet{labelcolor}{white!15!black}
\colorlet{ticklabelcolor}{white!45!black}
\colorlet{gridcolor}{white!85!black}

\colorlet{legendbordercolor}{white!100!black}

\pgfmathsetmacro{\groupplotsep}{17}
\pgfmathsetmacro{\timeplotheight}{2.8}
\pgfmathsetmacro{\timeplotwidth}{7.8}
\pgfplotsset{time-plot/.style={
		scale only axis,
		legend cell align={left},
		legend style={fill opacity=0, draw opacity=0, text opacity=1, draw=legendbordercolor, font=\scriptsize},
		tick align=inside,
		tick pos=left,
		xmajorgrids,
		ymajorgrids,
		grid style={gridcolor, line width=0.2pt, opacity=0.4},
		ticklabel style = {font=\scriptsize, ticklabelcolor,},
		y label style = {font=\footnotesize, labelcolor, rotate=-90, at=(ticklabel cs:1.1), anchor=west, inner sep=0pt},
		x label style = {font=\footnotesize, labelcolor},
		ytick style={draw=none},
		xtick style={draw=none},
		axis line style={draw=none},
		ylabel shift = -3 pt,
		xlabel near ticks,
}}

\pgfplotsset{position-plot/.style={
		scale only axis,
		axis equal image,
		legend cell align={left},
		legend style={fill opacity=0, draw opacity=1, text opacity=1, at={(0.03,0.97)}, anchor=north west, draw=legendbordercolor, font=\scriptsize},
		tick align=inside,
		grid style={gridcolor, line width=0.2pt, opacity=0.4},
		xmajorgrids,
		ymajorgrids,
		tick style={draw=none},
		ticklabel style = {font=\scriptsize, ticklabelcolor},
		y label style = {font=\footnotesize, labelcolor},
		x label style = {font=\footnotesize, labelcolor},
		xlabel = {x (m)},
		ylabel = {y (m)},
		axis line style = { draw = none },
}}

\pgfplotsset{soundfield-plot/.style={
		axis on top=true,
		scale only axis,
		axis equal image,
		axis line style = { draw = none },
		tick align = inside,
		tick pos=left,
		xtick style={color=black},
		ytick style={color=black},
		ticklabel style = {font=\scriptsize, ticklabelcolor},
		y label style = {font=\footnotesize, labelcolor},
		x label style = {font=\footnotesize, labelcolor},
		xlabel = {x (m)},
		ylabel = {y (m)},
		ylabel shift = -10 pt,
}}

\pgfplotsset{matrix-plot/.style={
		axis on top=true,
		scale only axis,
		axis equal image = true, 
		axis line style = { draw = none },
		tick align = inside,
		tick pos=left,
		xtick style={color=black},
		ytick style={color=black},
		ticklabel style = {font=\scriptsize, ticklabelcolor},
		y label style = {font=\footnotesize, labelcolor},
		x label style = {font=\footnotesize, labelcolor},
}}

\pgfplotsset{spectogram-plot/.style={
		axis on top=true,
		scale only axis,
		axis line style = { draw = none },
		tick align = inside,
		tick pos=left,
		xtick style={color=black},
		ytick style={color=black},
		ticklabel style = {font=\scriptsize, ticklabelcolor},
		y label style = {font=\footnotesize, labelcolor},
		x label style = {font=\footnotesize, labelcolor},
		ylabel shift = -10 pt,
		colorbar style={
			axis equal image = false,
			width=0.2*\pgfkeysvalueof{/pgfplots/parent axis width},
			height=0.8*\pgfkeysvalueof{/pgfplots/parent axis height},
			ylabel shift = 0 pt,
		},
}}

\pgfplotsset{
	micmarker/.style={
		semithick, 
		mark=*, 
		mark size=1.8, 
		only marks,
		draw=black,
		line width=0.5,
		mark options=solid,
}}

\pgfplotsset{
	srcmarker/.style={
		semithick, 
		mark=square*,
		mark size=3,
		only marks,
		draw=black, 
		line width=0.7,
		mark options=solid,
}}

\pgfmathsetmacro{\standardlinewidth}{1.5}
\pgfmathsetmacro{\offamount}{\standardlinewidth*1.7}
\pgfmathsetmacro{\standardlineopacity}{0.9}
\newcommand{\linejoinchoice}{round}

\pgfplotsset{
	line0/.style={
		color0, 
		opacity=\standardlineopacity, 
		line cap=round, 
		line width = \standardlinewidth pt, 
		mark options={solid},
		line join=\linejoinchoice,
}}

\pgfplotsset{
	line1/.style={
		color1, 
		opacity=\standardlineopacity, 
		dash pattern=on 1pt off \offamount pt,
		line cap=round,
		line width = \standardlinewidth pt, 
		mark options={solid},
		line join=\linejoinchoice,
}}

\pgfplotsset{
	line2/.style={
		color2, 
		opacity=\standardlineopacity, 
		dash pattern=on 3.5pt off \offamount pt, 
		line cap=round,
		line width = \standardlinewidth pt, 
		mark options={solid},
		line join=\linejoinchoice,
}}
\pgfplotsset{
	line3/.style={
		color3, 
		opacity=\standardlineopacity, 
		dash pattern=on 6pt off \offamount pt,
		line cap=round,
		line width = \standardlinewidth pt, 
		mark options={solid},
		line join=\linejoinchoice,
}}
\pgfplotsset{
	line4/.style={
		color4, 
		opacity=\standardlineopacity, 
		dash pattern=on 8pt off \offamount pt,
		line cap=round,
		line width = \standardlinewidth pt, 
		mark options={solid},
		line join=\linejoinchoice,
}}
\pgfplotsset{
	line5/.style={
		color5, 
		opacity=\standardlineopacity, 
		dash pattern=on 10pt off \offamount pt,
		line cap=round,
		line width = \standardlinewidth pt, 
		mark options={solid},
		line join=\linejoinchoice,
}}
\pgfplotsset{
	line6/.style={
		color7, 
		opacity=\standardlineopacity, 
		dash pattern=on 12pt off \offamount pt,
		line cap=round,
		line width = \standardlinewidth pt, 
		mark options={solid},
		line join=\linejoinchoice,
}}

\pgfplotsset{
	linegrad0/.style={
		color0!20!white, 
		opacity=\standardlineopacity, 
		line cap=round, 
		line width = \standardlinewidth + 2.5 pt, 
		mark options={solid},
		line join=\linejoinchoice,
}}

\pgfplotsset{
	linegrad1/.style={
		color0!45!white, 
		opacity=\standardlineopacity, 
		line cap=round, 
		line width = \standardlinewidth pt, 
		mark options={solid},
		line join=\linejoinchoice,
}}

\pgfplotsset{
	linegrad2/.style={
		color0!80!white, 
		opacity=\standardlineopacity, 
		line cap=round, 
		line width = \standardlinewidth pt, 
		mark options={solid},
		line join=\linejoinchoice,
}}

\pgfplotsset{
	linegrad3/.style={
		color0!85!black, 
		opacity=\standardlineopacity, 
		line cap=round, 
		line width = \standardlinewidth pt, 
		mark options={solid},
		line join=\linejoinchoice,
}}

\pgfplotsset{
	linegrad4/.style={
		color0!55!black, 
		opacity=\standardlineopacity, 
		line cap=round, 
		line width = \standardlinewidth pt, 
		mark options={solid},
		line join=\linejoinchoice,
}}

\pgfplotsset{
	linegrad5/.style={
		color0!10!black, 
		opacity=\standardlineopacity, 
		line cap=round, 
		line width = \standardlinewidth pt, 
		mark options={solid},
		line join=\linejoinchoice,
}}

\definecolor{color0}{rgb}{0.00392156862745098, 0.45098039215686275, 0.6980392156862745}
\definecolor{color1}{rgb}{0.8705882352941177, 0.5607843137254902, 0.0196078431372549}
\definecolor{color2}{rgb}{0.00784313725490196, 0.6196078431372549, 0.45098039215686275}
\definecolor{color3}{rgb}{0.8352941176470589, 0.3686274509803922, 0.0}
\definecolor{color4}{rgb}{0.8, 0.47058823529411764, 0.7372549019607844}
\definecolor{color5}{rgb}{0.792156862745098, 0.5686274509803921, 0.3803921568627451}
\definecolor{color6}{rgb}{0.984313725490196, 0.6862745098039216, 0.8941176470588236}
\definecolor{color7}{rgb}{0.5803921568627451, 0.5803921568627451, 0.5803921568627451}
\definecolor{color8}{rgb}{0.9254901960784314, 0.8823529411764706, 0.2}
\definecolor{color9}{rgb}{0.33725490196078434, 0.7058823529411765, 0.9137254901960784}

\colorlet{color0grad0}{color0!20!white}
\colorlet{color0grad1}{color0!45!white}
\colorlet{color0grad2}{color0!80!white}
\colorlet{color0grad3}{color0!85!black}
\colorlet{color0grad4}{color0!55!black}
\colorlet{color0grad5}{color0!10!black}

\colorlet{color3grad0}{color3!20!white}
\colorlet{color3grad1}{color3!45!white}
\colorlet{color3grad2}{color3!80!white}
\colorlet{color3grad3}{color3!85!black}
\colorlet{color3grad4}{color3!55!black}
\colorlet{color3grad5}{color3!10!black}

\usepackage{etoolbox}
\usepackage{siunitx}
\DeclareSIUnit\octave{oct}
\robustify\bfseries

\usepackage[acronym]{glossaries}
\newacronym{ces}{CES}{complex elliptically symmetric}
\newacronym{scm}{SCM}{sample covariance matrix}
\newacronym{dof}{DoF}{degrees of freedom}
\newacronym{cdf}{CDF}{cumulative distribution function}
\newacronym{pdf}{PDF}{probability density function}

\newacronym{dft}{DFT}{discrete Fourier transform}
\newacronym{fft}{FFT}{fast Fourier transform}

\newacronym{rir}{RIR}{room impulse response}
\newacronym{iir}{IIR}{infinite impulse response}
\newacronym{fir}{FIR}{finite impulse response}
\newacronym{ism}{ISM}{image-source method}

\newacronym{szc}{SZC}{sound zone control}
\newacronym{vast}{VAST}{variable span trade-off filter}
\newacronym{acc}{ACC}{acoustic contrast control}
\newacronym{pm}{PM}{pressure matching}

\newacronym{qos}{QoS}{quality of service}
\newacronym{fpi}{FPI}{fixed point iteration}

\newacronym{mse}{MSE}{mean square error}
\newacronym{nmse}{NMSE}{normalized mean square error}
\newacronym{ac}{AC}{acoustic contrast}
\newacronym{sd}{SD}{signal distortion}
\newacronym{re}{RE}{residual energy}
\newacronym{sinr}{SINR}{signal-to-interference-plus-noise ratio}
\newacronym{snr}{SNR}{signal-to-noise ratio}

\newacronym{evd}{EVD}{eigenvalue decomposition}
\newacronym{gevd}{GEVD}{generalized eigenvalue decomposition}
\newacronym{pevd}{PEVD}{polynomial eigenvalue decomposition}
\newacronym{pgevd}{PGEVD}{polynomial generalized eigenvalue decomposition}
\newacronym{svd}{SVD}{singular value decomposition}
\newacronym{mwf}{MWF}{multichannel Wiener filter}

\newacronym{ola}{OLA}{overlap-add}
\newacronym{ols}{OLS}{overlap-save}
\newacronym{wola}{WOLA}{weighted overlap-add}

\newacronym{lms}{LMS}{least mean squares}
\newacronym{nlms}{NLMS}{normalized least mean squares}
\newacronym{fxlms}{FxLMS}{filtered-x least mean squares}
\newacronym{rls}{RLS}{recursive least squares}

\newacronym{pseq}{PSEQ}{perfect sequence}
\newacronym{ops}{OPS}{orthogonal periodic sequence}
\newacronym{pps}{PPS}{perfect periodic sequence}

\newacronym{mnf}{MNF}{minimum noise fraction}
\newacronym{rkhs}{RKHS}{reproducing kernel Hilbert space}
\newacronym{krr}{KRR}{kernel ridge regression}
\newacronym{rff}{RFF}{random Fourier features}

\newacronym{psd}{PSD}{positive semi-definite}
\newacronym{sdr}{SDR}{semi-definite relaxation}
\newacronym{sdp}{SDP}{semi-definite program}
\newacronym{qcqp}{QCQP}{quadratically constrained quadratic program}

\newacronym{ot}{OT}{optimal transport}
\newacronym{pot}{POT}{partial optimal transport}
\newacronym{fgw}{FGW}{fused Gromov-Wasserstein}

\usepackage{dsfont}

\newcommand{\posmic}{\bm{m}}
\newcommand{\possrc}{\bm{s}}
\newcommand{\posism}{\bm{r}}

\newcommand{\srctime}{\tau}
\newcommand{\plan}{\bm{M}}
\newcommand{\weights}{\omega}

\newcommand{\injection}{\pi}

\title{Optimal transport of image sources for interpolation of room impulse responses with moving sources}
\name{Jesper Brunnström$^{\star}$\orcidlink{0000-0003-2946-1268}, Filip Elvander$^{\dagger}$\orcidlink{0000-0003-1857-2173}, Isabel Haasler$^{\star}$\orcidlink{0000-0002-2484-0181}\thanks{This work was supported by the Research Council of Finland under Grant 362787.}}
\address{$^{\star}$ Department of Information Technology, Uppsala University, Sweden \\ $^{\dagger}$ Department of Information and Communications Engineering, Aalto University, Finland}

\ninept 

\begin{document}
	\maketitle
	\begin{abstract}
    In geometrical acoustics, \acrfullpl{rir} can be represented by a set of image sources in free space. For a fixed source position, the image sources allow for computing \acrshortpl{rir} at arbitrary receiver positions. However, for a moving physical source, the image sources also move, making interpolation more difficult. In this paper we develop an interpolation method for image source positions of a moving source, given image source positions at the start and end of the trajectory. The method exploits the fact that each image source moves the same distance as the physical source. A statistical model is developed to derive cost functions and an appropriate dummy cost used in the proposed partial optimal transport (POT) approach. Through simulated experiments, POT using the proposed cost functions is shown to be effective compared to the alternatives. 


	\end{abstract}

	\section{Introduction}
    A \gls{rir} represents the acoustic path from a source to a receiver. Knowledge of \glspl{rir} is fundamental to a wide range of applications in audio signal processing, including speech enhancement and noise control \cite{zhangSurround2017, vorlanderAuralization2020, koyamaSpatial2021, naylorSpeech2010}. Therefore, measuring and interpolating \glspl{rir} represents a key problem in acoustic signal processing \cite{huangIdentification2006, uenoSound2025}.  A popular approach is to represent \glspl{rir} in a region by multiple sources in free space, making interpolation to other receiver positions straightforward \cite{antonelloRoom2017, tsunokuniSpatial2021, damianoCompressive2024}. 

    In geometrical acoustics \cite{polackPlaying1993, saviojaOverview2015}, the \gls{rir} can be represented by a set of image sources in free space, each corresponding to a sequence of specular reflections. This is referred to as the \gls{ism}, proposed in \cite{allenImage1979} and later extended to model the acoustics more accurately \cite{borishExtension1984, samarasingheSpherical2018, xuSimulating2024a, ewertComputationallyefficient2025}. The image source positions are independent of receiver position, and the model can therefore be used for \gls{rir} interpolation and acoustical analysis of the room \cite{sundstromOptimal2024}. However, image source positions depend on the position of the physical source, therefore an interpolation method is required for a moving source.  
 
    Without detailed knowledge of the wall geometry, which here is assumed unavailable, the image source positions must be obtained from collected sound data \cite{puomioLocating2021}. The visibility of an image source depends on both source and receiver position \cite{stephensonComparison1990}, meaning two measured sets of image sources will contain image sources not present in the other. It is therefore essential to find an assignment between two sets of image sources while discarding those without a good match, a task well suited to \gls{pot} \cite{benamouIterative2015, chapelPartial2020}.


    The proposed method differs from the \gls{pot} approach in \cite{geldertInterpolation2023} in that a moving source instead of a moving receiver is considered in this paper, which is the case where an interpolation method is crucial. In addition, well motivated cost functions and dummy costs are provided, which increase interpolation performance. The moving source case is considered using \gls{ot} in \cite{sundstromEstimation2024}, but without explicitly modelling image sources, which is the key to the proposed approach.

    To effectively interpolate, the structure of the image source geometry should be exploited. In particular, when the physical source moves, each image source moves the same distance in a direction depending on the unknown room geometry. In this paper, this structure is encoded into a statistical model, from which a maximum likelihood estimator for the assignment is obtained. This is used to derive cost functions for the \gls{pot} problem that exploit the geometric structure. In addition, the statistical model is used for a principled approach to determining which points should be discarded. 
   


    \section{Background}
    
    \subsection{Image-source method}
Consider a room $\Omega \subset \real^3$ with a physical source located at $\possrc \in \Omega$ and assume that the acoustic path from the source to a receiver at position $\posmic \in \Omega$ can be described by an \gls{rir} $h(\posmic, \possrc, t)$, at time $t$. Furthermore, assume that there exist $I \in \mathbb{N}$ image sources at positions $\posism_1, \dots, \posism_I$ with weights $\weights_1, \dots, \weights_I \in \real$ such that
    \begin{equation}
        h(\posmic, \possrc, t) = \sum_{i = 1}^{I} \frac{\weights_i}{4 \pi\lVert \posmic - \posism_i \rVert_2} \varphi \Bigl(t - \frac{\lVert \posmic - \posism_i \rVert_2}{c} \Bigr).
        \label{eq:rir-model}
    \end{equation}
That is, the contribution of each image source is a scaled and delayed impulse $\varphi$, where $c$ in \eqref{eq:rir-model} is the speed of sound.
   
    In the ideal case for infinite bandwidth, $h$ is a sum of time-domain Green's functions~\cite[Ch.7]{morseTheoretical1986}. In practice, the model is implemented in discrete time with bandlimited $\varphi$. As will be described in the next section, the position of each image source depends on the physical source position $\possrc$.

    \subsection{Position of image sources}\label{sec:image-source-placement}
Assume that the room $\Omega$ is a polyhedron with $W\in \naturals$ faces, each of which can be defined by a unit length normal $\bm{n} \in \real^3$ and a point on the face $\bm{p} \in \real^3$. For each face, a first-order image source is placed by mirroring the physical source in the corresponding face. In particular, for face $w$, the first-order image source is placed at
    \begin{equation}
        T_{w}(\possrc) = (\bm{I} - 2 \bm{n}_w \bm{n}_w^\transpose) \possrc + 2 \bm{n}_w \bm{n}_w^\transpose \bm{p}_w  = \bm{H}_{w} \possrc + \bm{b}_w, 
    \end{equation}
    where $\bm{H}_w$ is an orthogonal Householder matrix. Higher-order image sources are placed by reflecting lower order sources, starting from first-order. For example, the position of an $N$th order image source reflected through the sequence of faces $\mathcal{W} = w_1, w_2, \dots, w_N$ is 
    \begin{equation}
        T_{\mathcal{W}}(\possrc) = \bm{H}_{\mathcal{W}} \possrc + \bm{b}_{\mathcal{W}},
        \label{eq:general-ism-position}
    \end{equation}
    where $\bm{H}_{\mathcal{W}} = \bm{H}_{w_N} \cdots \bm{H}_{w_1}$ is still orthogonal, and $\bm{b}_{\mathcal{W}} \in \real^3$ depends only on the room geometry. Even though all image sources are placed according to \eqref{eq:general-ism-position}, not all image sources are necessarily visible from a given receiver position \cite{stephensonComparison1990}. Hence, only a subset will be simulated or measurable in practice. This also means that as a source or receiver moves, image sources can appear or disappear.

    \section{Problem statement}
    Consider a physical source moving along a linear trajectory from $\possrc_0$ to $\possrc_1$, i.e., any point on the trajectory can be described as $\possrc_\srctime = (1-\srctime)\possrc_0 + \srctime \possrc_1$ for $\srctime \in [0,1]$.  The positions and weights of the image sources up to order $N$ for each point $\srctime$ along the trajectory can be represented by a discrete non-negative measure $\mu_\srctime \in \mathcal{M}_+(\real^3)$, written explicitly as 
    \vspace{-0.2 cm}\begin{equation}
        \mu_\srctime = \sum_{i=1}^{I} \weights_i^{(\srctime)} \delta_{T_{\mathcal{W}_i}(\bm{s}_\srctime)},
    \end{equation}
    where $\bm{r}_i^{(\srctime)} = T_{\mathcal{W}_i}(\bm{s}_\srctime)$ is the image source position corresponding to the reflection sequence $\mathcal{W}_i$ for a physical source at $\bm{s}_\srctime$. The weights $\weights_i^{(\srctime)}$ depend on $\srctime$ due to potential occlusion. Throughout, we assume that $\weights_i^{(\srctime)} = 1$ for visible image sources and zero otherwise. Given $\mu_\srctime$, the \gls{rir} at any receiver position can be computed.
    
    The problem considered in this paper is to estimate $\mu_\srctime$ for all $\srctime \in [0,1]$ without explicit knowledge of the room geometry, given noisy estimates of $\mu_0$ and $\mu_1$. The \glspl{rir} may then be interpolated by convolving with the appropriate basis functions in \eqref{eq:rir-model}. 
    
    We assume access to estimates\footnote{Using a microphone array, the image source positions can be estimated using, e.g., the method in \cite{puomioLocating2021}} of $\mu_0$ and $\mu_1$, denoted $\nu_0$ and $\nu_1$, corresponding to known physical source positions $\possrc_0$ and $\possrc_1$, respectively. Specifically, 
\begin{equation}
        \nu_0 = \sum_{i=1}^{I_0} \delta_{\bm{x}_i} \;,\quad \nu_1 = \sum_{j=1}^{I_1} \delta_{\bm{y}_j}.
        \label{eq:data-model}
    \end{equation}
In general, $I_0 \neq I_1$ due to visibility changes or missed detections, and the measured image sources are obtained in no particular order. The estimated image source positions $\bm{x}_i$ and $\bm{y}_j$ are modeled as
    \begin{equation}
        \begin{aligned}
            \bm{x}_i = \bm{r}_i^{(0)} + \bm{\epsilon}_{i}^{(0)} \;,\quad
            \bm{y}_{j} = \bm{r}_{\injection(j)}^{(1)} + \bm{\epsilon}_{j}^{(1)},
        \end{aligned}
        \label{eq:data-model-positions}
    \end{equation}
where $\bm{\epsilon}_{i}^{(0)}$ and $\bm{\epsilon}_{j}^{(1)}$ represent estimation error. Without loss of generality, the image sources are indexed so that $\bm{x}_i$ is a measurement of $\bm{r}_i^{(0)}$ for $i = 1, \dots, I_0$, and $i > I_0$ corresponds to image sources not detected at $\possrc_0$. The correspondence between $\bm{x}_i$ and $\bm{y}_j$ is described by the unknown injection $\injection: \{1,\ldots,I_{1}\}\to \{1,\ldots,I\}$. This means $\bm{x}_i$ and $\bm{y}_j$ are measurements of the same image source when $\injection(j) = i \leq I_0$, which occurs for $I_{01} = \lvert \{ j : \injection(j) \leq I_0 \} \rvert$ pairs. When all detected image sources are visible at both endpoints, i.e. $I_0 = I_1 = I_{01}$, the map $\injection$ reduces to a permutation of $\{1, \dots, I_{01}\}$. Herein, we propose to recover the common subset and the correspondence on it using a \gls{pot} formulation.

    \section{Algorithm}\label{sec:algorithm}
    \subsection{Partial optimal transport}
    The assignment and interpolation problem can be solved using \gls{pot}. \Gls{pot} is equivalent to balanced \gls{ot} where mass can be transported to a dummy point~\cite{chapelPartial2020}, and can be written \cite{baiSliced2023}
    \begin{equation}
        \begin{aligned}
            \minimize_{\substack{\plan \geq 0 \\ \bm{u}, \bm{v} \geq 0}} &\quad\langle \bm{C}, \plan \rangle_F + \xi \bm{1}^\transpose (\bm{u} + \bm{v})  \\
            \text{subject to }  &\quad\plan \bm{1} + \bm{u} = \bm{1} \\
            &\quad\plan^\transpose \bm{1} + \bm{v} = \bm{1}. \\
        \end{aligned}
        \label{eq:optimization-problem}
    \end{equation}
    The cost matrix $\bm{C} \in \real^{I_0 \times I_1}$ is defined as $\bm{C}_{ij} = c(\bm{x}_i, \bm{y}_j)$, for some appropriate ground cost $c : \real^3 \times \real^3 \rightarrow \real_{\geq 0}$. The transport plan is $\plan \in \real^{I_0 \times I_1}$, and $\bm{u} \in \real^{I_0}, \bm{v}\in \real^{I_1}$ are the transport plans to the dummy points. The parameter $\xi \geq 0$ represents the cost of transporting mass to a dummy point. The optimization problem \eqref{eq:optimization-problem} is a linear program, and can be solved using convex optimization \cite{diamondCVXPY2016}. There always exists an optimal plan $\plan$ which is a partial one-to-one matching, taking only values $\{0, 1\}$, and each point is paired with at most one point of the other set \cite[Th.~4.1]{baiSliced2023}, which means $\plan$ well serves the purpose of estimating the assignment $\injection$. 



    \subsection{Interpolation} \label{sec:interpolation}
    Displacement interpolation can be well motivated by considering the geometric structure of the image sources. Specifically, when the physical source moves, the position of any image source moves the same distance. This can be shown explicitly as  
    \begin{equation}
    \begin{aligned}
        &\lVert \posism_i^{(0)} - \posism_i^{(1)} \rVert_2 = \lVert T_{\mathcal{W}_i}(\possrc_0) - T_{\mathcal{W}_i}(\possrc_1) \rVert_2 \\
        &= \lVert \bm{H}_{\mathcal{W}_i} (\possrc_0 -\possrc_1) \rVert_2 = \lVert \possrc_0 -\possrc_1 \rVert_2,
    \end{aligned}
    \end{equation}
    where the last equality uses the orthogonality of $\bm{H}_{\mathcal{W}_i}$. More specifically, as the physical source moves along $\possrc_\srctime$, the image sources also move along linear trajectories $\posism_i^{(\srctime)} = (1-\srctime) \posism_i^{(0)} + \srctime \posism_i^{(1)}$.

    The proposed interpolation scheme is therefore displacement interpolation \cite{mccannConvexity1997}, which is computed as 
    \begin{equation}
        \hat{\mu}_\srctime \! =  \!\sum_{i=1}^{I_0} \sum_{j=1}^{I_1} \! M_{ij} \delta_{\bm{x}_{i}(1-\srctime) + \srctime \bm{y}_j}\! + \sum_{i=1}^{I_0} (1-\srctime) u_i \delta_{\bm{x}_i}\! + \sum_{j=1}^{I_1} \srctime v_j \delta_{\bm{y}_j}
        \label{eq:interpolated-measure}
    \end{equation}
    after solving \eqref{eq:optimization-problem}. The first term represents transport between image sources present in both $\nu_0$ and $\nu_1$, and the last two terms fade out points that are determined to only exist in either $\nu_0$ or $\nu_1$. Note that the interpolated measure \eqref{eq:interpolated-measure} will contain weights taking any value in $[0,1]$, despite $\nu_0$ and $\nu_1$ only consisting of unit weights. Under ideal conditions, where the transport plan $\plan$ exactly recovers the assignment $\injection$, no mass is moved to a dummy point, and $\bm{x}_i$ and $\bm{y}_j$ contain no noise, then the proposed interpolation scheme is exact.

    \section{Statistical model}\label{sec:statistical-model}
    \begin{figure}
        \centering
        \includegraphics{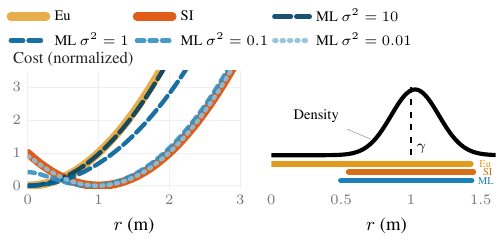}
        \vspace{-0.2cm}
        \caption{The cost functions viewed as functions of $r=\lVert \bm{x} - \bm{y}\rVert_2$, normalized such that their minimum values are 0 (left), and the density of $r$ together with the accepted regions for $\alpha = 0.01$ (right). \vspace{-0.4cm}}
        \label{fig:cost-illustration}
    \end{figure}

    The proposed approach described in Section~\ref{sec:algorithm} requires a cost function and a choice of dummy cost $\xi$. In this section, we introduce a statistical model to derive effective choices for both. Throughout this section, it is assumed that $\injection$ is a permutation, that is $I_0 = I_1 = I_{01}$. 

    The true residual of two measured image source positions \eqref{eq:data-model-positions} with correct assignment is 
\begin{equation}
    \bm{\rho}_i^{*} = \bm{y}_{\injection^{-1}(i)} - \bm{x}_i
    = \bm{H}_{\mathcal{W}_i} (\possrc_1 - \possrc_0)
      + \bm{\epsilon}^{(1)}_{\injection^{-1}(i)} - \bm{\epsilon}^{(0)}_{i} .
    \label{eq:true-residual-model}
\end{equation}
    The room geometry is considered to be unknown, which means $\bm{H}_{\mathcal{W}_i}$ is also not known. The term $\bm{H}_{\mathcal{W}_i} (\possrc_1 - \possrc_0)$ is a vector of known length $\gamma$ but unknown direction. Therefore, the residual $\bm{\rho}_i \in \real^3$ is modelled as 
    \vspace{-0.3cm}\begin{equation}
        \bm{\rho}_i = \gamma \bm{d}_i + \bm{\epsilon}_i,
    \end{equation}
    where $\gamma = \lVert \possrc_0 - \possrc_1 \rVert_2$, and $\bm{d}_i \sim \mathcal{U}(\unitsphere)$ is sampled with a uniform distribution on the unit sphere. The estimation error is modelled as independent Gaussian noise, $\bm{\epsilon}_i^{(0)} \sim \mathcal{N}(0, \sigma_0^2 \bm{I})$ and $\bm{\epsilon}_i^{(1)} \sim \mathcal{N}(0, \sigma_1^2 \bm{I})$, so the noise can be expressed as a combined noise vector $\bm{\epsilon}_i = \bm{\epsilon}_{\injection^{-1}(i)}^{(1)} - \bm{\epsilon}_i^{(0)} \sim \mathcal{N}(0, \sigma^2\bm{I})$ where $\sigma^2 = \sigma^2_0 + \sigma^2_1$. 





    \subsection{Assignment maximum likelihood estimator}
    According to the proposed statistical model, the distribution of the residual $\bm{\rho}_i$ given $\bm{d}_i$ is the shifted Gaussian $\bm{\rho}_i \vert \bm{d}_i \sim \mathcal{N}(\gamma\bm{d}_i, \sigma^2 \bm{I})$. The unconditional distribution can be obtained by marginalizing over the nuisance parameter $\bm{d}_i$, giving
    \begin{equation}
        p(\bm{\rho}) = \frac{1}{(2 \pi \sigma^2)^{3/2}} e^{\bigl(-\frac{r^2 + \gamma^2}{2\sigma^2} \bigr)} \sinhc(\frac{\gamma r}{\sigma^2}),
    \end{equation}
    where $\sinhc(x) = \frac{\sinh(x)}{x}$, and $r = \lVert \bm{\rho} \rVert_{2}$. 

    Viewing image source positions of $\nu_1$ as $\bm{y}_{\injection^{-1}(i)} = \bm{\rho}_i + \bm{x}_i$ for deterministic $\bm{x}_i$, the likelihood of the data for an assignment $\injection$ is $\mathcal{L}(\injection) = \prod_{i=1}^{I_{01}} p(\bm{y}_{\injection^{-1}(i)} - \bm{x}_i)$. This expression is obtained under the assumption that each $\bm{\rho}_i$ is independent, motivated by a lack of knowledge about the room. Any knowledge of room geometry or reflection order implies a dependence between the positions. Using the shorthand $r_{ij} = \lVert \bm{x}_i - \bm{y}_j\rVert_{2}$, the assignment $\injection$ that maximizes the data likelihood can be obtained by solving
    \vspace{-0.2cm}\begin{equation}
    \begin{aligned}
        &\argmax_{\injection} \mathcal{L}(\injection) = \argmin_{\injection} \sum_{i=1}^{I_{01}} \bigl[- \log p(\bm{y}_{\injection^{-1}(i)} - \bm{x}_i) \bigr] \\
        &= \argmin_{\injection} \sum_{i=1}^{I_{01}} r_{i\injection^{-1}(i)}^2 - 2 \sigma^2\log \sinhc(\frac{\gamma r_{i\injection^{-1}(i)}}{\sigma^2}).
    \end{aligned}\vspace{-0.2cm}
    \label{eq:maximum-likelihood}
    \end{equation}
        
    \subsection{Ground cost}
    Solving the maximum likelihood problem \eqref{eq:maximum-likelihood} is equivalent to solving a balanced \gls{ot} problem with the cost
    \begin{equation}
        c_{\text{ML}}(\bm{x}_i, \bm{y}_j) = r_{ij}^2 + \gamma^2 - 2 \sigma^2\log \sinhc(\frac{\gamma r_{ij}}{\sigma^2}).
        \label{eq:data-likelihood}
    \end{equation}
    Therefore, \eqref{eq:data-likelihood} is proposed as a statistically motivated cost function for the algorithm \eqref{eq:optimization-problem}. The constant $\gamma^2$ in \eqref{eq:data-likelihood} is added to obtain a nonnegative cost function.

    Two other costs are considered. First, the commonly used Euclidean cost is considered, defined as
    \begin{equation}
    \begin{aligned}
        c_{\text{Eu}}(\bm{x}_i, \bm{y}_j)  &= r_{ij}^2. 
    \end{aligned}
    \end{equation}
    The cost $c_{\text{Eu}}$ is asymptotically equivalent to $c_{\text{ML}}$ when $\sigma \gg \gamma$. In particular, when $\gamma \leq \sqrt{3} \sigma$, both $c_{\text{ML}}$ and  $c_{\text{Eu}}$ have the same minimum, when $r_{ij} = 0$. As a simpler alternative to $c_{\text{ML}}$, we also propose a source-informed cost $c_{\text{SI}}$ which takes the distance $\gamma$ into account, defined as
    \begin{equation}
    \begin{aligned}
        c_{\text{SI}}(\bm{x}_i, \bm{y}_j) &= (r_{ij} - \gamma)^2.  \vspace{-0.2cm}
        \end{aligned}
    \end{equation}
    The cost $c_{\text{ML}}$ is asymptotically equivalent to $c_{\text{SI}}$ when $\gamma \gg \sigma$, which is illustrated in Fig.~\ref{fig:cost-illustration}.

    

    \subsection{Choosing dummy cost parameter $\xi$}
    For good performance, $\xi$ has to be chosen appropriately, which depends on the cost function. In \eqref{eq:optimization-problem}, only mass between points satisfying $c(\bm{x}_i, \bm{y}_j) \leq 2\xi$ will be transported \cite[Lemma 3.2]{baiSliced2023}. Therefore, the parameter $\alpha$ is introduced, which is the probability that a matched pair falls outside the accepted region, i.e., that $c(\bm{x}_i, \bm{y}_{\injection^{-1}(i)}) > 2 \xi$. The parameter $\alpha$ is interpretable independently of the cost function. 

    

    Each cost function can be written in terms of $r = \lVert \bm{x} - \bm{y} \rVert_2$, which according to the statistical model follows $r \sim \sigma \chi_3(\frac{\gamma}{\sigma})$, where $\chi_3$ is a non-central chi distribution with 3 degrees of freedom. Therefore, the cumulative distribution function of $c(\bm{x}, \bm{y})$ for each cost is known, which can be used to solve $\text{Prob}(c(\bm{x},\bm{y}) > 2 \xi) = \alpha$ numerically \cite[Ch.~3-5]{brentAlgorithms2013}. The density of $r$ together with the accepted regions for the different cost functions for the same $\alpha$ is shown in Fig.~\ref{fig:cost-illustration}.

    \begin{figure*}
        \centering
        \includegraphics{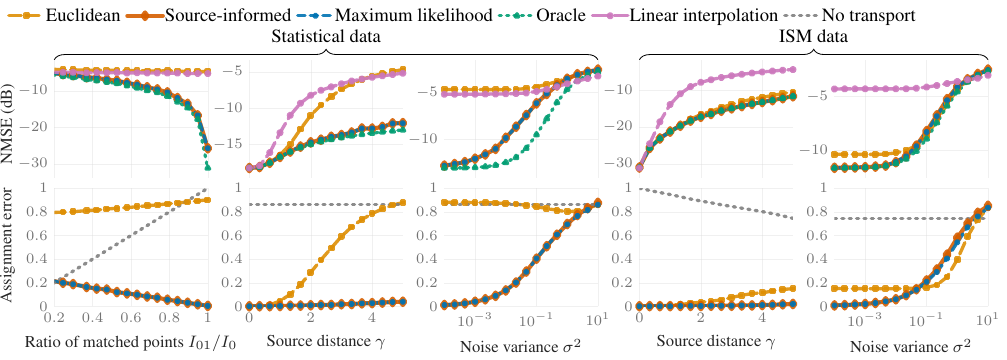}\vspace{-0.2cm}
        \caption{The result in terms of NMSE and assignment error for all considered experiment scenarios. The assignment error for oracle is always 0, and therefore not shown. \vspace{-0.5 cm}}
        \label{fig:rir-error}
    \end{figure*}

    \section{Experimental validation}
    \subsection{Error metrics}
    The first performance metric is the assignment error, describing how well $\injection$ is estimated. The assignment $\injection$ defines an ideal plan $\plan^\star \in \real^{I_0 \times I_1}$ which is $\plan^\star_{ij} = 1$ if $\injection(j) = i$ and $0$ otherwise. The ideal dummy transport plan is $\bm{u}^\star = \bm{1} - \plan^\star \bm{1}$. The assignment error is then defined as 
    \begin{equation}
        \text{Assignment error } = (\sum_{ij} \lvert \plan_{ij} - \plan^\star_{ij} \rvert + \sum_{i} \lvert \bm{u}_i - \bm{u}^\star_i \rvert) / 2 I_0. \vspace{-0.2cm}
    \end{equation}

    The second performance metric is the \gls{nmse} of the estimated \glspl{rir}. Let $\bm{h}_\mu \in L^2(\real \rightarrow \real^M)$ denote the \glspl{rir} generated by a measure $\mu$ through \eqref{eq:rir-model} at the receiver positions $\posmic_1, \dots, \posmic_M \in \Omega$. The \gls{nmse} for a source at position $\possrc_\srctime$ is defined as
    \vspace{-0.3cm}\begin{equation}
        \text{NMSE}_\srctime = \frac{\lVert \bm{h}_{\mu_\srctime} - \bm{h}_{\hat{\mu}_\srctime}
        \rVert_{L^2}^2}{\lVert \bm{h}_{\mu_\srctime} \rVert_{L^2}^2} 
        \label{eq:nmse}
    \end{equation}
    where $\lVert \bm{h}_{\mu_\srctime} \rVert_{L^2}^2 = \sum_{m=1}^M \int_{\real} h_{\mu_\srctime}(\bm{m}_m, \bm{s}_\srctime, t)^2\,dt$. The \gls{nmse} can be evaluated without reconstructing the \glspl{rir} in discrete time by noting that the inner product that induces the norm $\lVert \cdot \rVert_{L^2}$ can be written in the form  $\langle \bm{h}_\mu, \bm{h}_{\mu'} \rangle_{L^2} = \sum_{i,j} \weights_i \weights'_j k(\posism_i, \posism'_j)$ for two measures $\mu = \sum_i \weights_i \delta_{\posism_i}$ and $\mu' = \sum_j \weights'_j \delta_{\posism'_j}$. For an ideal bandlimited impulse $\varphi(t) = 2B \sinc (2 \pi B t)$, the kernel function is
    \vspace{-0.2cm}\begin{equation}
        k(\posism, \posism') = \sum_{m=1}^{M} \frac{2B \sinc(2 \pi B \Delta_m)}{(4\pi)^2
        \lVert \posmic_m - \posism\rVert_2 \lVert \posmic_m - \posism' \rVert_2},
        \label{eq:nmse-kernel}
    \end{equation}
    with $\Delta_m = (\lVert \posmic_m - \posism' \rVert_2 - \lVert \posmic_m - \posism \rVert_2)/c$. Finally, the \gls{nmse} can be aggregated along the trajectory by approximating $\text{NMSE} = \int_{0}^1 \text{NMSE}_\srctime \,d\srctime$ with the trapezoid rule. Note that the metric is ill-posed at infinite bandwidth, where any non-zero position error leads to the maximum \gls{nmse} for that image source.

    

    \subsection{Methods}
    The proposed method using the cost functions $c_{\text{ML}}$, $c_{\text{SI}}$, and $c_{\text{Eu}}$, is referred to as \textit{maximum likelihood}, \textit{source-informed}, and \textit{Euclidean} respectively. The following methods are considered for comparison.

    \textit{Oracle} uses the ground truth assignment $\injection$ and performs interpolation according to Section~\ref{sec:interpolation} using the noisy position measurements. The method represents a ceiling for the methods using displacement interpolation, and indicates how much error is caused by measurement noise and the fading of unmatched image sources. 
    

    \textit{Linear interpolation} combines the measures linearly as $\hat{\mu}_\srctime = (1-\srctime) \nu_0 + \srctime \nu_1$. This is equivalent to applying linear interpolation between \glspl{rir} as $\bm{h}_{\hat{\mu}_\srctime} = \srctime \bm{h}_{\nu_1} + (1-\srctime)\bm{h}_{\nu_0}$.

    \textit{No transport} is a baseline for the assignment error, which sets $\bm{M} = 0$, and transports all mass to the dummy points. It is also indicative of how much mass should be transported by an ideal plan. 

    \subsection{Data}
    Two types of data are considered in the experiments. 

    \textit{Statistical data} is generated using the statistical model described in Section~\ref{sec:statistical-model}. The true image source positions $\bm{r}_i^{(0)}$ are sampled with uniform distribution over a cube of volume $V I_{0}$, where $V = \qty{120}{\meter^3}$, and $I_0 = I_1 = 57$, $I_{01} = 49$. The $M=16$ receivers are placed randomly with uniform distribution in a cube of volume $V$ that is centered in the larger cube. Each unmatched point changes visibility at a random point along the trajectory $\srctime_i \sim \mathcal{U}([0,1])$.


    \textit{ISM data} is generated as follows. With equal probability, the number of faces is randomly selected $W \in \{6, \dots, 8\}$. The walls are defined by a random convex polygon with $W-2$ corners \cite{valtrProbability1995}, with parallel floor and ceiling. The physical source position $\possrc_0$ is random with uniform distribution in the room, and $\possrc_1$ is randomly chosen $\gamma$ distance from the initial point with uniform distribution, both positions are placed at least \qty{0.5}{\meter} from the closest boundary. Image sources up to order $N=3$ are generated~\cite{scheiblerPyroomacoustics2018}. The $M=16$ receivers are placed randomly with uniform distribution in the room.

    \subsection{Results}
    All results are presented in Fig.~\ref{fig:rir-error}, where the mean result from 256 random trials is presented. The bandwidth is chosen as $B = \qty{250}{\hertz}$, and the rejection probability as $\alpha = 0.001$. Unless explicitly varied, $\sigma^2 = 10^{-3}$ and $\gamma = 5$.

    First, the proportion of image sources detected in both $\nu_0$ and $\nu_1$ was varied for statistical data. When this proportion is small, the \gls{nmse} is similar between all methods, despite source-informed and maximum likelihood attaining much lower assignment error. This indicates a performance floor for the \gls{nmse} caused by mismatch between the fading of image sources in \eqref{eq:interpolated-measure} and their true behavior. 

    Then, the \gls{nmse} and assignment error are computed for different values of $\gamma$ and $\sigma$, using both types of data. For statistical data, Euclidean and linear interpolation perform significantly worse compared to source-informed and maximum likelihood, which almost attain the oracle performance. For ISM data, the performance difference is smaller, with Euclidean only being slightly worse than maximum likelihood and source-informed. Note that source-informed and maximum likelihood perform indistinguishably from the oracle unless the noise variance is high. All \gls{pot} methods perform significantly better compared to linear interpolation, except at very high noise levels. 
    
    In the high-noise for ISM data, Euclidean performs better than source-informed and maximum likelihood, which indicates that the statistical model could be extended to capture more features of the ISM data. The results indicate that source-informed is the overall pragmatic choice, providing similar performance to maximum likelihood, while being simpler, as only the dummy cost $\xi$ depends on $\sigma$ and not the cost itself, so $\xi$ may also be set heuristically.

    \section{Conclusion}
    In this paper, an interpolation method for image sources based on \gls{pot} has been developed. By introducing a statistical model, appropriate cost functions could be derived, informed by the maximum likelihood estimator of the assignment between the data points. Through simulated experiments using both statistical data and data from the \gls{ism} for random rooms, the proposed \gls{pot} approach is shown to be more accurate than linear interpolation, with maximum likelihood and source-informed costs being particularly effective.

	\bibliographystyle{IEEEbib_compact}
	\bibliography{abbrev, refs}
	
\end{document}